# A Study on the Electrode Protection System of an 100t Ultra High Power Electric Furnace

UnChol Choe, YongChol Sin, CholGuk Yun

**Kim Il Sung** University,

Pyongyang,

Democratic People's Republic of Korea

**Abstract**

In this paper a method to design the electrode protection system using arc current and voltage and pressure inside the hydraulic cylinder of electrode lift device of an 100t ultra high power furnace (UHP furnace) and confident factor(CF) and Lyapnov function, and then we validated our method through the field experimentation.

## 1. Introduction

One of main instruments making steel, an UHP furnace has the low productive cost and very high productivity.

The electrode protection system to prevent the non-normal conditions such as the electrode broking which occurs during steel-melting is very important for the safe operation and saving electrode.

The previous works considered about control of melting in UHP are classified into approaches using the mathematical modeling of plant and approaches using soft-computing.

In the work [4] they proposed the electrode lift control method of the mode settling the arc current in which the on-off control is combined with the PID control, in the work [5] they proposed the parameter identification method and the advanced electrode lift PID control algorithm using least square method and pole-placement adaptive algorithm.

In the work [6] they proposed a reference model adaptive control using partly linear model instead of non-linear model and in the works[1,2,8,12] they framed the intelligent control system using soft-computing technology and then considered the electrode controlling issues.

In the previous works above, the core control value is arc current or voltage or impedance or arc length.

The necessary problem for electrode control and establishing the protective system of electrode is a model representing dynamics according to control operation [7~10].

The design theory and method about hybrid intelligent system aimed to multi-functions allow to stand the defaults of individual computing and to quantify the complicated events exactly [13,14].

In the work[15], they introduced the concept of confident factor into uncertainty inference and assessed the inference results with the value of confident factor .

Lyapnov function used in somedeciplines such as automatic control theory reflects the dynamic changes quantitatively to assessed stability of a dynamic system.

They didn't study to use confident factor and Lyapnov function together in predictive control or predictive control used soft computing.

That's why in the classical predictive control they don't need uncertainty inference because they use its mathematical model and the research on model for inference dynamic change viewed from change of energy is insufficient in the predictive control using soft computing.

In this paper we use arc current and the 2$^{nd}$ voltage and confident factor and Lyapnov function being used in inference to assess the change of melting process of 100t UHP and proposed the protective system for preventing electrode broking and validated through the introducing experimentation.

## 2. The principle of electrode protection during steel-melting

The principle diagram of the electrode lift control system in UHP is as figure 1.

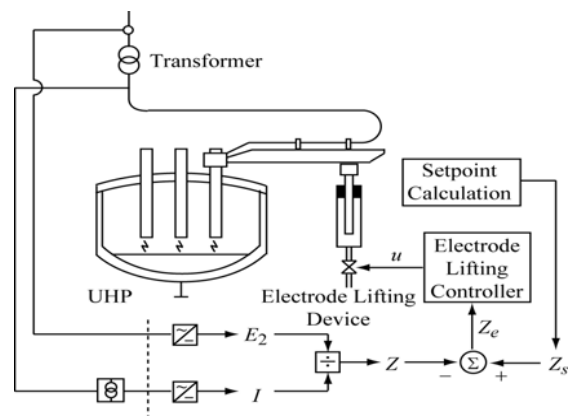

Figure 1. The principle diagram of the electrode lift control system in UHP

The electrode lift automatic control system is used to control the electrode lift automatically to settle arc length constantly in order to settle arc current and arc voltage and so on about their set points.

In this paper, arc current and the 2$^{nd}$ phase voltage and deviation of arc current and pressure inside hydraulic cylinder for electrode lift are selected as factors representing dynamic change of UHP.

And we spill the electrode lift control range into arc-generating range and normal range about set current and over-current range where arc current is over set current to be able to interrupt the normal operation and the dangerous current range.

From factors representing dynamic change of UHP and the electrode lift control ranges, the protection system is framed as follow.

① A protection loop to prevent accident when arc current is dangerous.
② A protection loop to prevent accident when arc current is more than over current.
③ A loop to identify non-metal by pressure change inside the hydraulic cylinder for electrode lift.
④ A Loop to identify charge collapse when voltage is low.

⑤ Predicting charge collapse using intelligent method.

## 3. A protection loop by means of arc current and pressure inside hydraulic device for electrode lift.

When arc current is over the dangerous current, maxim load current of furnace transformer, this causes an electrical accident in electrode lift device and power system.

And the frequent operation at lower current than dangerous current causes the serious oscillating of electrode that an electrode accident may occur.

The 1st and 2nd protection loops are realized by setting of the dangerous current and over current and rules setting the out of electrode lift controller.

The setting of over current and dangerous current and control out for electrode lift is done considering the maxim load current of furnace transformer whose capacity is 55,000KVA.

The set point of arc current in electrode lift control is from 35 000A to 45 000A.

$I_s = 35,000A \sim 45,000A$.

When arc current is more than dangerous current, we must lift the electrode fastest,

$I_a \geq I_{danger} = (I_s + 15,000A) \Rightarrow$ to lift the electrode fastest.

where $I_a$ is arc current and $I_{danger}$ is the dangerous current.

When the arc current is over current, we lift electrode at half of maximum velocity,

$I_a \geq I_{over} = (I_s + 7,500A) \Rightarrow$ to lift electrode at half of maximum velocity,

where $I_{over}$ is bias of over-current.

If arc current suspends about over current more than 1s, we lift electrode fastest.

The 3rd protection loop prevents the electrode broking when electrode meets non-metal.

When electrode presses the non-metal during electrode control, electrode is forced physically and at this time hydraulic pressure inside hydraulic cylinder lowers.

Therefore if the pressure inside electrode lift cylinder is lower than 0.9 times of normal pressure for 1.5s, we agree that electrode meets non-metal.

## 4. Identification of charge collapse and electrode protection by prediction.

The 4th and 5th protection loops prevent the electrode broking by physical and electrical impact when charge collapses into the hole been formed during melting stage.

### 1) Identification of charge collapse by low-voltage and electrode lift control.

Phase voltage $E_2$ is the vector sum of voltage at reactance $U_X$ and voltage at resistance $U_R$.

$$E_2 = U_X + U_R \qquad (1)$$
$$U_X = I_a X \qquad (2)$$

$$U_R = I_a(R_a + R_f) \tag{3}$$

Here $R_a, R_f$ are arc resistances and system resistances respectively.

If measured the 2$^{nd}$ phase voltage be close to $U_X$, $R_a$ decreases and it means that electrode met charge, namely the charge collapses.

We agree that the charge collapse when the following condition folds.

$$E_2 \leq 1.2 \cdot I_a X \Rightarrow \text{ charge collapses}$$

Then we must lift electrode fastest for 2s.

## 2) Prediction of charge collapse using intelligent method and electrode protection.

Fuzzy rule $R_j$ whose conclusion part is a constant is of the form,

$$\begin{aligned} r_j : &\text{if } (x_1 \text{ is } A_{j1}) \text{ and } \cdots \text{ and } (x_k \text{ is } A_{jk}) \text{ and } \cdots \text{ and } (x_n \text{ is } A_{jn}) \\ &\text{then } (y_{j1} \text{ is } a_{j1}) \text{ and } \cdots \text{ and } (y_{jk} \text{ is } a_{jk}) \text{ and } \cdots \text{ and } (y_{jp} \text{ is } a_{jp}) \end{aligned} \tag{5}$$

where $x_k (k = \overline{1,n})$ is the input variable of primary part of $j$ ($j = \overline{1,m}$) th fuzzy rule, $A_{jk}$ is the membership function for $x_k (k = \overline{1,n})$ in $j$ ($j = \overline{1,m}$) th fuzzy rule, $y_{jk} (k = \overline{1,p})$ is the output variable of the primary part in $j$ ($j = \overline{1,m}$) th fuzzy rule, $a_{jk}$ is the constant for $y_{jk} (k = \overline{1,p})$ in $j$ ($j = \overline{1,m}$) th fuzzy rule.

If we define the fuzzy membership functions as Gaussian function, it is possible to be correspond the $j$ ($j = \overline{1,m}$) th rule to a RBF neural network.

In this paper we represent the dynamic change as the quantified assessment rule using the confident factor and the change of Lyapnov function and consider a problem making a predictive model with RBF NN network reflecting assessment rule.

**(1). Making assessment rules using confident factor and Lyapnov function.**

We define the language set assessing the dynamic change of plant at a moment as {to change well, to change badly, to not change}.

In this paper we introduced the confident factor being used in inference uncertainty and Lyapnov function representing varying energy of dynamic characteristics.

We select error of arc current from set current, $e \in R^{3\times 1}$ and error deviation $\dot{e} \in R^{3\times 1}$ at 2 moments previous and now.

We selected the deviation of Lyapnov function to be predicted at t moment, $\dot{V}_t$ and the confident factor about dynamic change, $CF_t$.

At this time we made rule for assessment about dynamic change of plant as follows,

$$r_j : if \ x_{11} = \mu_{11} \ and \ \cdots \ x_{1n} = \mu_{1n} \ and \ x_{21} = \mu_{21} \cdots and \ x_{2n} = \mu_{2n}$$
$$then \ y_{j1} = a_1 \ and \ y_{j2} = a_2 \ and \ y_{j3} = a_3 \quad (6)$$

where $r_j$ is the $j$th rule, $x_{ki}$ is a variable, $\mu_{ki}$ is the property function value according $x_{ki}$, $a_k$ is the property function value about the conclusion $y_{jk}, k = \overline{1,3}$ to represent the predictive state about dynamic change in rules.

The range of index is $(k = \overline{1,2}, \ i = \overline{1,n})$ and at the predict moment $a_1$ denotes bigness of orientation change of dynamics, $a_2$ denotes the confident factor of orientation change, $a_3$ denotes the change orientation at next moment.

We selected Lyapnov function as follows,

$$V = X_1^T P X_1 \quad (7)$$
$$P > 0$$

where $X_1 = (x_{11}, \cdots x_{1n})^T = (e(t-1), \cdots e(t-n))^T$, $X_1 \in R^{n \times 1}$.

We denote the variable representing the Lyapnov function at last moment as $y_0 = a_0$ and predicted about change of dynamic characteristics as follows.

Table2. qualitative results of change orientation of dynamic characteristics by qualitative inference

| Sign combination of $a_0, a_1$ | change orientation of dynamic characteristics |
|---|---|
| $a_0 \leq 0, a_1 < 0$ | To change to fine orientation |
| $a_0 \geq 0, a_1 > 0$ | To change to bad orientation |
| $a_0 = 0, a_1 = 0$ | No change |
| $a_0 \leq 0, a_1 > 0$ | To oscillate and change to bad orientation |
| $a_0 \geq 0, a_1 < 0$ | To oscillate to change to good orientation |

The 2$^{nd}$ variable of conclusion part in the rule allows to predict the change of dynamic characteristics with confident factor $CF$.

Therefore we make the 6$^{th}$ assessment rule as follows,

$$R_1 : if \ x_{11} = \mu_{11} \ \cdots and \ x_{1n} = \mu_{1n} \ and \ x_{21} = \mu_{21} \cdots$$
$$\cdots and \ x_{2n} = \mu_{2n} \ then \ y_{j1} = a_1 \ and \ y_{j2} = a_2 \ and \ y_{j3} = a_3 \quad (8)$$

$$R_2 : if \ x_{11} = \mu_{11} \ \cdots and \ x_{1n} = \mu_{1n} \ and \ x_{21} = \mu_{21} \cdots$$
$$\cdots and \ x_{2n} = \mu_{2n} \ then \ y_{j1} = a_1 \ and \ y_{j2} = a_2 \ and \ y_{j3} = a_3 \quad (9)$$
$$R_3 : if \ x_{11} = \mu_{11} \ \cdots and \ x_{1n} = \mu_{1n} \ and \ x_{21} = \mu_{21} \cdots$$
$$\cdots and \ x_{2n} = \mu_{2n} \ then \ y_{j1} = a_1 \ and \ y_{j2} = a_2 \ and \ y_{j3} = a_3 \quad (10)$$

where $R_i, i = \overline{1,3}$ denotes the set of rules formed with several rules.

We represent $a_0 = \dfrac{\dot{V}}{\dot{V}_{max}}$ and select the property functions of conditional variables of the jth rules in the rule set $R_i, i = \overline{1,3}$ as follows,

$$\mu_{1i} = \exp[-\dfrac{(x_{1i} - \theta_{j1})^2}{2\sigma_{j1}^2}] \quad (11)$$

$$\mu_{2i} = \exp[-\dfrac{(x_{2i} - \theta_{j2})^2}{2\sigma_{j2}^2}] \quad (12)$$

where $\theta_{ji}$ denotes the mean and $\sigma_{ji}$ denotes the band of Gaussian function.

The property function value of each variable in conclusion part of rule is as follows.

$$a_1 = \dfrac{\dot{V}_t}{\dot{V}_{max}} \quad (13)$$

$$a_2 = CF = \dfrac{1}{3} \cdot (|a_1| + \tilde{\mu}_1 + \tilde{\mu}_2) \quad (14)$$

$$\tilde{\mu}_1 = \dfrac{1}{n} \sum_{i=1}^{n} \mu_{1i} \ , \quad \tilde{\mu}_2 = \dfrac{1}{n} \sum_{i=1}^{n} \mu_{2i}$$

The value of $a_3$ representing the change orientation of dynamic characteristics is as follows.

if $a_0 \leq 0, a_1 < 0$, $a_3 = 1$. if $a_0 \geq 0, a_1 > 0$, $a_3 = -1$

if $a_0 = 0, a_1 = 0$, $a_3 = 0$. if $a_0 \leq 0, a_1 > 0$, $a_3 = -0.5$

if $a_0 \geq 0, a_1 < 0$, $a_3 = 0.5$

Equations above allow to predict the change of dynamic characteristics with the valued value in "how much" according to CF.

Because we must to analyze lots of data, we manipulate the issue of making the predictive model by RBF NN network reflected the assessment rules.

**(2) Construction of RBF NN network reflected assessment rules.**

$x_{ki} \ (k = \overline{1,2}, \ i = \overline{1,n})$, the variables of primary part are selected as input variables of NN network.

And $a_1$ and $a_2$, $a_3$, the property function values of the out of conclusion part $y_i, i = \overline{1,3}$ are selected as outs of NN network.

The fuzzy rules are expressed as equations (8)~(10) and the membership functions of primary part are expressed as equations (11)~(12) and then the firing strength of $j, (j=\overline{1,M})$ th rule is as follows,

$$w_j = \prod_{k=1}^{2} w_{jk} \qquad (15)$$

$$w_{j1} = \prod_{i=1}^{n} \exp[-\frac{(x_{1i}-\theta_{j1})^2}{2\sigma_{j1}^2}], \quad w_{j2} = \prod_{i=1}^{n} \exp[-\frac{(x_{2i}-\theta_{j2})^2}{2\sigma_{j2}^2}]$$

where the inference result at the $j$ th rule of rule set with M rules is as follows,

$$y_k = \frac{\sum_{j=1}^{M} w_j \cdot y_{jk}}{\sum_{j=1}^{M} w_j} = \sum_{j=1}^{M} \tilde{w}_j y_{jk}, \quad (k=\overline{1,3}) \qquad (16)$$

where $\tilde{w}_j = \dfrac{w_j}{\sum_{j=1}^{M} w_j}$.

The structure of RBF NN network combined with the valued rules is as follows.

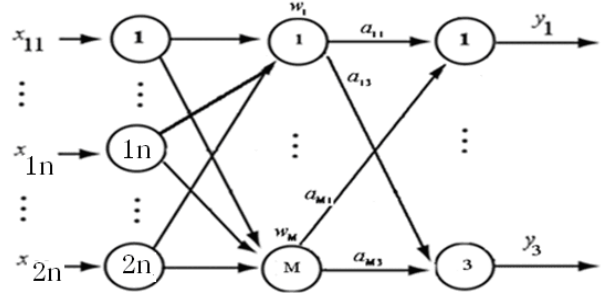

Figure2. The structure of RBF NN network combined with the valued rules

The error of out of the NN network from target about Nth training sample is as follows,

$$E_n = \frac{1}{2}\sum_{k=1}^{3}(T_k - y_k)^2 \qquad (17)$$

where $y_k = \dfrac{\sum_{j=1}^{M} w_j \cdot y_{jk}}{\sum_{j=1}^{M} w_j} = \sum_{j=1}^{M} \tilde{w}_j y_{jk}$ and M is number of cells in the hidden layer.

We update the parameters in the form,

$$U_j^{new} = U_j^{old} + \eta \Delta U_j \qquad (18)$$

where $\eta$ is the training rate, $\Delta U_i$ is the modification of parameter.

The modification of parameter are as follows.

$$\Delta\theta_{ji} = -(T_k - y_k) \cdot \tilde{w}_j \cdot (a_k - y_k) \cdot \frac{(x_j - \theta_{ji})}{\sigma_{ji}^2} \quad (19)$$

$$\Delta\sigma_{ji} = -(T_k - y_k) \cdot \tilde{w}_j \cdot (a_k - y_k) \cdot \frac{(x_j - \theta_{ji})}{\sigma_{ji}^3} \quad (20)$$

$$\Delta a_{jk} = -(T_k - y_k) \cdot \tilde{w}_j \cdot \quad (21)$$

So we use the calculation results of the predictive model with RBF NN network to protect electrode.

## 5. Introducing experiments

### 1) Simulation results by MATLAB and analysis

We show a curve predicting the charge collapse by CF which is out of the NN network as figure 3.

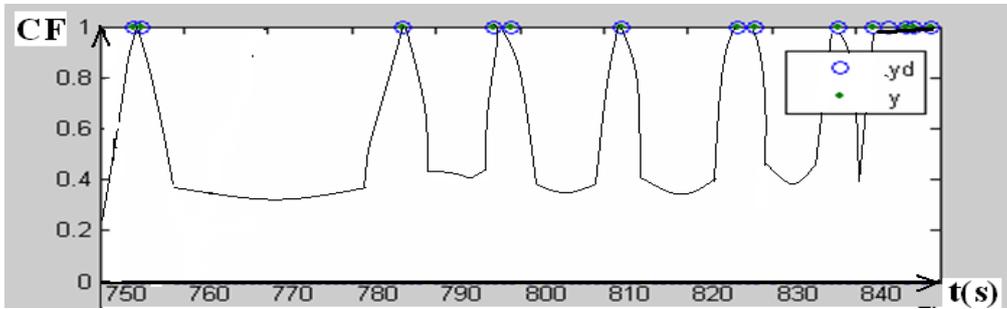

Figure 3. a curve predicting charge collapse

In this figure, the points whose CF equals 1 predict the charge collapse and the circles are the real charge collapse. There are charge collapses after 13 minutes and big charge collapse after 13 minutes 40 seconds and the accuracy of the predicting collapse is about 90%.

### 2) Introductions and effectiveness

We use the fuzzy self-tuning PI control in the electrode lift automatic control.

From the simulation above we made the identification loop of charge collapse and rules setting control out.

If the change rate of variance of arc current increase in the same sign, $a_1$ is big and the CF of orientation change $a_2$ is over 0.6, we see the change orientation at the next moment $a_3 = -1$ and alert the charge collapse and lift the electrode which changes fastest at 30% of maxim lift velocity.

[Step 1]: to save the derivations of arc currents of 5 moments previous from now.

[Step 2]: if each derivation above is bigger than 300A, go to Step 3, else go to Step 5.

[Step 3]: summing the arc current derivations at 2 moments previous from now and name it Sum1, and summing the arc current derivations from 3 moments previous to 5 moments previous and name it sum2.

[Step 4]: if Sum1-Sum2>6000, alert the charge collapse and lift an electrode which changes

fastest at 30% of the maximum velocity.

[Step 5]: end

In 100t UHP furnace we examine 218 charges, among them we examined 98 charges with the Fuzzy self-tuning PI control using our electrode protection system and we examined 120 charges with the old method.

As result we shared 2.7kg/t of electrode per 1t of steel.

## Conclusion

In this paper we studied to use arc current and the 2nd phase voltage and the pressure inside hydraulic system for electrode lift and the confident factor and Lyapnov function assessing the change of dynamic characteristics of 100t UHP and proposed the electrode protection system in order to prevent the electrode broking.

We also validated our method through the field experimentation in 100t UHP.